\documentstyle[12pt]{article}
\begin{document}

\title{Physics of Finance}
\author{
Kirill Ilinski
\thanks{E-mail: kni@th.ph.bham.ac.uk}
\\ [1cm]
{\small\it IPhys Group, CAPE,
14-th line of Vasilievskii's Island, 29}\\
{\small\it St-Petersburg, 199178, Russian Federation}
\\ [0.3cm]
{\small\it School of Physics and Space Research,
University of Birmingham,} \\
{\small\it Edgbaston B15 2TT, Birmingham, United Kingdom}
\\ [0.3cm]
{\small\it Theor. Department, Institute for Spectroscopy,
Troitsk,\ \ \ \ \ \ \ \ \ \ \ } \\
{\small\it  \ \ \   Moscow region, 142092, Russian Federation }
}

\vspace{3cm}

\date{ }

\vspace{1cm}
 
\maketitle

\vskip 1 true cm

{\bf
We give a brief introduction to the Gauge Theory of 
Arbitrage~\cite{KI}. 
Treating a calculation of net present values (NPV) 
and currencies
exchanges as a parallel transport in some fibre bundle,
we give geometrical interpretation 
of the interest rate, 
exchange rates
and prices of securities as a proper 
connection components. This allows us
to map the theory of capital market onto 
the theory of quantized gauge field
interacted with a money flow field.
The gauge transformations of the matter field
 correspond to a
dilatation (redefinition) of security units 
which effect is eliminated by 
a proper
tune of the connection. The curvature tensor for 
the connection consists
of the excess returns to the risk-free interest 
rate for the local arbitrage
operation. Free quantum gauge theory is equivalent 
to the assumption
about the log-normal walks of assets prices.
In general case the consideration maps the capital 
market onto QED, i.e.
quantum
system of particles with positive (securities) 
and negative ("debts")
charges which interact with each other through 
electromagnetic field
(gauge field of the arbitrage). In the case of 
a local virtual
arbitrage opportunity money flows in the 
region of configuration
space (money poor in the profitable 
security) while "debts" try
to escape from the region. Entering positive
 charges and leaving
negative ones screen up the profitable 
fluctuation and restore the
equilibrium in the region where there is 
no arbitrage opportunity  any
more, i.e. speculators washed out the 
arbitrage opportunity.
} \\
\thispagestyle{empty}
\setcounter{page}{0}
\newpage

\section{Informal sketch}
There is a common belief that nature 
(and a human society, in particular) 
could be described in some unified terms 
and concepts of physics may find 
their use here. The present paper follows 
the last point and tries
to draw parallels between
the theory of financial markets and 
quantum gauge theory.
Since it is difficult to be equally comprehensible for
both physicists and economists we 
give the introduction which is
basically intuitive and designed to explain 
main logic of the
consideration. Further sections are more 
mathematical and
use notions of quantum field theory.

\subsection{Net present value as a parallel transport}

First of all let us recall what the 
NPV method is. The NPV investment
method works on the simple, but 
fundamental, principle that money have a
time value. This time value has to 
be taken into account through
so-called discounting process.

To elucidate the idea we use here 
an easy example from
Ref.~\cite{Lumby}:  "Suppose you were 
made an offer: if you pay 500
pounds now, you will immediately 
receive 200 French Francs, 200 Japanese
Yen and 200 German Marks. How  would you 
go about deciding whether the
offer was worthwhile? What you certainly 
would not do is to say: I have
to give up 500 pieces of paper and, in 
return, I will get 600 pieces of
paper. As I end up with 100 more pieces 
of paper, the deal is worthwhile!
Instead, you would recognize that in order 
to evaluate the offer, you
have to convert all the different currencies 
to a common currency, and
then undertake the comparison" (please, 
do not consider this as a Euro
propaganda!). In the same way that money in
 different currencies cannot
be  compared directly, but first has to 
be converted to a common
currency, money in the same currency, but 
which occur at different points
in time cannot be compared directly, but 
must first converted to a common
point of time. This reflects the time 
value of money.

Intuitively it is clear that given the choice 
between $\L$100
now or $\L$100 in one year time, most 
people would take the
$\L$100 now since the money could be 
placed on deposit (risk free
investment) at some interest $r$ rate. 
Then, in one year time the $\L$100
turn itself into $\L (1+r)$100 instead of keeping 
$\L$100 only. Therefore
$r$ - the interest rate - represents 
the time value of money.

Skipping here some details (difference between simple and
compound interest rate, continuous compounding, 
flat and effective
interest rates~\cite{Blake}) we now are 
ready to formulate what the NPV
is: if an amount of money $F$ is to be 
received in $T$ years' time, the
Present Value of that amount ($NPV(F)$) 
is the sum of money $P$
(principal) which, if invested today, 
would generate the compound amount
$F$ in $T$ years' time:
$$
NPV(F) \equiv P = \frac{F}{(1+r)^T} \ .
$$
The interest rate involved in this calculation 
is known as the discount
rate and the term $(1+r)^{-T}$ is known as 
T-year discount factor $D_T$:
\begin{equation}
D_T = (1+r)^{-T} \ .
\label{DT}
\end{equation}
In a similar way, to calculate the present value 
of a stream of payments,
the above formula is applied to each individual 
payment and the resulting
individual present values are then summed.
So, the NPV method states that if the NPV of an 
investment project has
zero or positive value, the company should 
invest in the project; if it
has negative value, it should not invest. 
There is also some use of the
NPV for comparative analysis of several 
projects but we do not stop
for the details (see~\cite{Lumby}).

What is really important here for our goals is the following
geometrical interpretation: discounting procedure 
plays a role of a
"parallel transport" of a money amount through a 
time  (though in fixed
currency). The discounting factor (\ref{DT}) is 
then an element of a
structural group of a fibre bundle (which still 
has to be defined) and
the discount rate coincides with the 
time component of the connection
vector field.  
Let us remind that in the differential geometry 
the connection field is 
responsable
for pulling a fibre element from on point of a base 
space to another.
Moreover, it is obvious that the "space" components of
the connection has something to do with the exchange 
rates (see the
example above).  Indeed, the exchange rates or 
prices are responsible for
converting money in different
currencies or different securities (read points of 
discrete "space") to the
same currency (point of the space) at a fixed moment of time.
So, they
can be interpreted as elements of the structural group 
which "transport"
the money in "space" directions and are space analogues 
of the discount
factor.  Summing up, we see that there is some analogue 
between
elements of a fibre bundle picture with some connection 
field and a
capital market.  We make the analogue precise below.

\subsection{Arbitrage as a curvature of the connection}

Next keyword is an
arbitrage. What then the arbitrage is? 
Basically it means "to get
something from nothing" and a free lunch after 
all. More strict
definition states the arbitrage as an operational 
opportunity to make a
risk-free profit~\cite{Wilmott} with a rate of 
return higher
than the risk-free interest rate accured on 
deposit (formalized version
can be found in \cite{Duffie}).

The arbitrage appears in the theory when
we consider a curvature of the connection. 
In more details, a rate of
excess return for an elementary arbitrage 
operation (a difference between
rate of return for the operation and the 
risk-free interest rate) is an
element of curvature tenzor calculated from the connection. 
It can be understood keeping in mind that a
 curvature tensor elements are 
related to
a difference between two results of 
infinitasimal parallel transports 
performed in
different order. In financial terms it means that 
the curvature tensor elements
measure a difference in gains accured from two 
financial operations with the 
same
initial and final points or, in other words, a 
gain from an arbitrage operation.

In a certain sense,
the rate of excess return for an 
elementary arbitrage operation is an
analogue of the electromagnetic field. In an absense 
of any uncertanty (or,
in other words, in an absense of walks of prices, 
exchange and interest
rates) the only state is realised is the state of 
zero arbitrage.However,
if we place the uncertenty in the game, prices and 
the rates move and
some virtual arbitrage possibilities to get more 
than less appear. Therefore we 
can
say that
the uncertanty play the same role in the developing theory as the
quantization did for the quantum gauge theory.

{\bf Money flows as matter fields}

Last principle ingredient to enters the theory is  
"matter" fields
which interact through the connection. It is clear 
now that the "matter" fields 
are money flows fields which have to be gauged 
by the connection.
Indeed, we started the introduction of the 
concept with the example of NPV 
method
which shows how an amount of money units changes 
while the payment is "pulled" 
in
time or one currency is converted to another.

Dilatations of money units (which do not change a 
real wealth) play a
role of gauge transformation which eliminates the effect of the
dilatation by a proper tune of the connection 
(interest rate, exchange
rates, prices and so on) exactly as the 
Fisher formula does for the
real interest rate in the case of an 
inflation~\cite{Lumby}.  {\it The symmetry
of the real wealth to a local dilatation of money 
units (security splits
and the like) is the gauge symmetry of the theory}.

Following a formal analogy with $U(1)$ gauge theory 
(electrodynamics) case,
we can say that an amount of a certain currency units 
at a particular time 
moment
is analogues to a value of a phase cross section of the 
$U(1)\times M$ fibre 
bundle
at some space-time point. If we want to compare values 
of the $U(1)\times M$ 
cross
section at different points of the space-time we have 
to perform a parallel
transport of the phase from one point of the base to 
another exactly as we have 
to
convert one currency to another or discount money in 
the financial setting.

A theory may contain several types of the "matter" 
fields which may differ, for
example, by a sign of the connection term as it is  for
 positive and negative
charges in the electrodynamics. In the financial stage 
it means different
preferances of investors. Thus, in the paper we deal with cash 
flows and
debts flows which behave differently in the same gauge field: cash tries to 
maximize
it-self while debts try to minimize them-selves. This is 
equivalent to a 
behaviour
of positive and negative charges in the same electric 
field where they move in
different directions.

Investor's strategy is not always optimal. It is due to partially
incomplete information in hands, choice procedure, 
partially because of investor (or
manager) internal objectives. It means that the money 
flows are not
certain and fluctuates in the same manner as
the prices and rates do. So this requires a statistical 
description of the money
flows which, once again, returns us to an effective quantization 
of the theory.

\subsection{All together is QED-like system}

Collecting this all together, we are ready to map the 
capital market to a
system of particles with positive (securities) and negative ("debts")
charges which interact with each other through 
electromagnetic field
(gauge field of the arbitrage). In the case of a local virtual
arbitrage opportunity cash-debts flows in the region of 
configuration
space (money poor in the profitable security) while "debts" try
to escape from the region. Entering positive charges and leaving
negative ones screen up the profitable fluctuation and 
restore the
equilibrium so that  there is no arbitrage opportunity any
more. It means speculators wash up the arbitrage.

Taking into account the uncertainty mentioned above, 
we come to the quantum 
field
theory with the gauge field and "matter" fields 
of opposite charges. At this 
point
standard machinery of quantum field theory 
may be applied to obtain 
distribution
functions of the interest/exchange rates and 
cash-debts flows correlators (which
are essential for the response of the system). 
It may answer questions about
dynamical response of a financial market, dynamical 
portfolio theory and to be 
an approach to option pricing and other problems.

Last point to note in the subsectioin is a principal 
lattice nature of the 
capital
market theory, since there always exists a natural 
minimal time interval 
(transaction time) and the
"space" is a graph. This remove usual problems of 
the quatum field theory which 
has to deal with divergences due to contiuous 
character of its space-time base. 

\subsection{Basic assumptions}

In the previous subsections we discussed the main 
assumption (or postulate) 
which is
a {\it local gauge invariance} with the dilatation 
gauge group. It can be shown 
that the postulate essentially dictates dynamical rules 
of the theory.
However, there is a number of other assumptions which we want 
to list here.
For the convinience we devide all assumptions into two parts.
First of them deals with the prices and rates , i.e. 
contains assumptions
concerning the connection field.
\begin{enumerate}
\item Exchange rates, prices of various securities and 
interest rate
fluctuate and it bears local arbitrage opportunities.
\item The most probable configuration of the random 
connection is the
configuration with a minimal absolute value of the excess 
rate of return.
\item Arbitrage opportunities, in absense of money flows, are
uncorrelated in different space-time points, i.e. the events 
are statistically 
independant.
\item We suppose exponential  distribution of the rate of 
return on a
local arbitrage operation. Its characteristics, in general, 
depend on both
"space" and time coordinates. However, for sake of simplicity,
 below we do not
consider the details.
\item When a continuous limit will be taken, we assume all required
smothness properties of relevant objects.
\end{enumerate}

Second part of main assumtions consists
of ones about a behavior of cash-debts flows i.e. the "matter" fields.
\begin{enumerate}
\item We assume the perfect capital market enviroment, i.e. there are
always a possibility to place money on a deposit and 
to borrow without
any restrictions and at the same interest rate \cite{note1}.
\item There are transaction costs. Their presense 
is not just unimportant
complication. The transaction costs play a role of inertia for the
cash-debts flows and stabilize the system.
\item Investors are (limited) rational and try to 
maximize their gain from the
securities, minimizing debts at the same time.
\item Exactly as the rates and prices may fluctuate 
and are uncertain,
flow trajectories fluctuate around the most
profitable trajectory ("classical trajectory"). 
This reflects the fact
that investors not always behave optimally.
\end{enumerate}
Below we repeat the assumptions in more strict form
which is useful for further formalisation of the theory.

\section{Formal constructions}

In this section we formalize the previous consideration. 
More precise, we give a
description of the relevant fibre bundles, we construct 
the parallel transport  rules using for this elements of 
the structural 
group and give an interpretation of the parallel transport
operators. The corresponded curvature is also defined 
and it is shown that
it is equal to the rate of excess return on the 
elementary plaquette
arbitrage operation. This opens a way to a 
construction of the
dynamics of the parallel transport factors what gives 
a lattice gauge
theory formulation. The procedure of the dynamics 
construction is
repeated then for the case of "matter" fields which
 represent cash-debts flows. 

\subsection{Fibre bundle construction}

It is well-known that many important concepts in physics can be
interpreted in terms of the geometry of fibre bundles~\cite{EGH}.
Maxwell's theory of electromagnetism and Yang-Mills theories are
essentially theories of the connections on principal 
bundles with a given
gauge group $G$ as the fibre. Einstein's theory of 
gravitation deals with
the Levi- Civita connection on the frame bundle of 
the space- time
manifold. In the section we show how the construction 
of fibre bundle can
also be applied to describe a framework  to develop 
the capital market
theory.

{\bf Construction of the base}

Now we have to construct a base of the fibre bundle.
Let us order the complete set of assets (which we want to
analyze) and label them by numbers from $0$ to $N$. 
This set can be
represented by $N$ (asset) points on a 2- dimensional plane (the
dimension is a matter of convinience and can be choosen arbitrary). 
To
add the time in the construction we attach a copy of 
$Z$ -lattice (i.e. set
of all integer number $\{..., -1, 0, 1, 2, ... \}$) 
to the each asset
point. We use discretized time since anyway 
there is a natural time
step and all real trades happen discretely.
All together this gives the prebase set $L_0 = \{1,2,...,N\}
\times Z$.

Next step in the construction is to define a 
{\it connectivity} of the
prebase. To do this we start with an introduction of a 
matrix of links
$\Gamma : L_0\times L_0\rightarrow \{ 0, \pm 1 \}$ which is 
defined by the
following rule for any $x\equiv (i,n)\in L_0$ and
$y\equiv (k,m)\in L_0$:
$\Gamma (x,y) =0$ except for
\begin{enumerate}
\item $i=k$ and $n=m - 1$ accepting that the 
$i$-th security exists at
the $n$-th moment and this moment is not an expiration date for the
security;
\item $n=m - 1$ and at $n$-th moment of time the $i$-th 
asset can be exchanged
on some quantity of $k$-th asset at some rate (we assume 
that the trasaction 
takes 
one unit of time).
\end{enumerate}
In latter situations $\Gamma (x,y) = 1 = -\Gamma (y,x)$.

Using the matrix $\Gamma (.,.)$ we define a {\it curve} 
$\gamma (x,y)$ in
$L_0$ which links two points $x,y\in L_0$.  We call the set 
$\gamma (x,y)
\equiv \{x_{j}\}_{j=1}^{j=p}$ a curve in $L_0$ with ends at points
$x,y\in L_0$ and $p-1$ segments if $x=x_{1}$, $x_{p}=y$, 
$\forall x_{j}\in
L_0$ and
$$
\Gamma (x_{j},x_{j+1}) = \pm 1 \qquad \mbox{for} \qquad \forall
j=1,...,p-1 \ .
$$

The whole $L_0$ can be devided into a set of 
connected components. A
connected component is a maximal set of elements 
of $L_0$ which can be
linked by some curve for any pair of elements.
The base $L$ is defined now as the
connected component contained US dollars at, say, 
15.30 of 17-th of June
1997. This complete the construction of the base of fibre bundle.

{\bf Structural group}

The structural group $G$ to be used below is a group 
of dilatations. The corresponding irreducable representation
 is the following:
the group $G$ is a group of maps $g$ of $R_+ \equiv ]0,+\infty )$ 
to
$R_+$, which act as a muliplication of any $x\in R_+$ on some
positive constant $\lambda (g)\in R_+$:
$$
g(x) = \lambda (g)\cdot x \ .
$$
Transition functions of a fibre bundle with the structure group
correspond below to various swap rates, exchange rates, 
discount factors
for assets.

{\bf Fibres}

In the paper we use fibre bundles with the following fibres $F$:
\begin{enumerate}
\item $F=G$, i.e. the fibre coincides with the structure group. The
corresponding fibre bundle is called Principle fibre bundle $E_P$.
A gauge theory in the fibre bundle in the next section 
corresponds to random
walks of prices and rates.
\item $F=R_+$.
This fibre bundle  will be important to describe cash-debts flows. 
Indeed,
a {\it cross section} (or simply a {\it section}) $s$ 
(a rule which
assigns a preferred point $s(x)$ on each fibre to each 
point $x\in L$ of
the base) of the fibre bundle is a "matter" field.
In the context $s(x\equiv (i,m))$
gives a number of units
of $i$-th assert at the moment of time $m$.
\end{enumerate}
Actions and the corresponding functional integrals will be 
written in terms of 
cross
sections of the fibre bundles. The main property of the objects 
(actions and
measures of the integrations) will be local gauge invariance, 
i.e. independance 
on
a local action of the structural group.

The fibre bundle $E$ we use below is trivial, i.e. 
$E=L\times F$ and we do 
not stop for the definition of projections.  
Construction of the fibre bundles for the simple stock exchange, 
FX-market and
financial derivatives can be found in Ref\cite{KI}.

\subsection{Parallel transport, curvature and arbitrage}

A connection is a rule of the parallel transport of an element
of a fibre from one point ($x$) of a base to another point ($y$).
It means that an operator of the parallel transport 
along the curve
$\gamma$, $U(\gamma ): F_x \rightarrow F_y$ is an element of the
structural
group of the fibre bundle~\cite{EGH}. Since we do not 
deal with continuous case and
restrict our-selves on a lattice formulation, we do not need 
to introduce
a vector -field of the connection but rather have to 
use elements of the
structural group $G$. By the definition, an operator of the 
parallel
transport along a curve $\gamma$, $U(\gamma )$, defined 
as a product of
operators of the parallel transport along the links
which constitute the curve $\gamma$:
$$
U(\gamma ) = \prod_{i=1}^{p-1} U(x_i,x_{i+1})\ , 
\qquad \gamma \equiv
\{x_i\}^{i=1}_{i=p-1}\ , \qquad x_1=x, \quad x_p=y \ .
$$
It means that we need to define only the parallel 
transport operators
along elementary links. Since $U(\gamma)=U^{-1}(\gamma^{-1})$, this
restrict us on a definition of those 
along an elementary links with a positive connectivity.
Summing up, the rules of the parallel transport in the 
fibre bundles are
completely defined by a set of the parallel transport operators 
along
elementary links with a positive connectivity. The definition 
of the set
is equivalent to a definition of the parallel transport in 
the fibre
bundle.

Since in subsection 2.1 the connectivity was defined by a 
possibility of
assert movements in "space" and time, it allows us to give an
interpretation of the parallel transport. In the subsection 
two principle
kinds of links with positive connectivity were defined. First one
connects two points $(i,n)$ and $(i,n+1)$ and represents a 
deposition the
$i$-th assert for one unit of time. This deposition then 
results in a
multiplication of the number of assert units by an 
interest factor
(or internal rate of return factor) calculated as:
$$
U((i,n),(i,n+1)) = e^{r_i \Delta} \in G \ ,
$$
where $\Delta$ is a time unit and $r_i$ is an appropriate 
rate of return
for the $i-th$ assert. In the continuous limit $r_i$ becomes 
a time
component of the corresponding connection vector-field at 
the point
$(i,\Delta n)$.

By the same way the parallel transport operator is defined 
for the second
kind of the elementary links, i.e. links between $(i,n)$ and 
$(k,n+1)$ if
there is a possibility to change at $n$-th moment a unit of 
$i$-th assert
on $S^{i,k}_{n}$ units of $k$-th assert:
$$
U((i,n),(k,n+1)) = S^{i,k}_{n} \in G \ .
$$
In general, an operator of the parallel transport along a curve is a
multiplier by which a number of assert units is multiplied as a 
result of
an operation represented by the curve.

Results of parallel transports along two different curves with the 
same
boundary points are not equal for a generic set of the parallel 
transport 
operators. A measure of the difference is a curvature tensor $F$. 
Its
elements are equal to resulting change in multiplier due to a 
parallel
transport along a loop around an infinitasimal elementary 
plaquette with
all nonzero links in the base $L$:
$$
F_{\mbox{plaquette}\rightarrow 0} = \prod_{m} U_{m} - 1 \ .
$$
 The index $m$ runs over all plaquette links, $\{U_{m}\}$ are
corresponding parallel transport operators and an agreement about
an orientation is implied.

Now we show that the elements of the curvature tensor are,
 in fact, an
excess returns on the operation corresponded to a plaquette.
Since elements of the curvature tensor are local quantities, it
is sufficient to consider
an elementary plaquette on a "space"-time base graph. Let us, 
for example, 
consider two different assets (we will call them for 
the moment share and cash) which can be
exchanged to each other with some exchange rate $S_i$ (one share is
exchanged on $S_i$ units of cash) at some moment $T_i$, and 
the reverse
rate
(cash to share) is $S^{-1}_i$. We suppose that there exists a
transaction time $\Delta$ and this $\Delta$ is taken as a time unit.
So the exchange rates $S_i$
are quoted on a set of the equidistant times: $\{T_i\}_{i=1}^N,
T_{i+1}-T_i =\Delta$. Interest rate for cash is $r_1$ so that
between two subsequent times $T_i$ and $T_{i+1}$ the volume 
of cash is
increased by factor $e^{r_1 \Delta}$. The shares are 
characterized by
a rate $r_2$.
As we will show latter, due to the
gauge invariance we can fix $r_1$ to be
risk free interest rate and $r_2$ related
to the average rate of return of the share.

Let us consider an elementary (arbitrage) operation between two
subsequent times $T_i$ and
$T_{i+2}$. 
There are two possibilities for an investor,
who posseses a cash unit at the moment $T_i$, to get shares
by the moment $T_{i+2}$.
The first one is to put cash on a bank
deposit
with the interest rate $r_1$ at the moment $T_i$, 
withdraw money back at
the moment
$T_{i+2}$ and buy shares for price $S_{i+1}$ each. 
In this way
investor gets $e^{r_1 \Delta}S_{i+1}^{-1}$ shares at the 
moment $T_{i+2}$
for each unit of cash he had at the moment $T_i$. 
The second way is to 
buy the
shares for price $S_i$ each at the moment $T_i$. Then, 
at the moment
$T_{i+2}$ investor will have $S_i^{-1} e^{r_2 \Delta}$ 
shares for each
unit
of cash at the moment $T_i$.  
If these two numbers ($ e^{r_1\Delta}S_{i+1}^{-1}$ and
$S_i^{-1} e^{r_2 \Delta}$) are not equal then there is a 
possibility for
an arbitrage. Indeed, suppose that $e^{r_1\Delta}S_{i+1}^{-1} <
S_i^{-1} e^{r_2 \Delta}$,
then at the moment $T_i$ an arbitrager can borrow one unit of cash, 
buy
$S_i^{-1}$ shares and get $S^{-1}_i e^{r_2 \Delta}S_{i+1}$ 
units of cash
from selling shares at the moment $T_{i+1}$. The value of this cash
discounted to the moment $T_i$  is
$S^{-1}_i e^{r_2 \Delta}S_{i+1} e^{-r_1 \Delta}>1$. 
It means that $S^{-1}_i e^{r_2 \Delta}S_{i+1} e^{-r_1 \Delta}-1$ 
is an
arbitrage excess return on the operation. In the other hand, 
as we have
shown above, this represent lattice regularisation of an 
element of the
curvature tensor along the plaquette.
If $e^{r_1 \Delta}S_{i+1}^{-1}>S_i^{-1} e^{r_2 \Delta}$ 
then arbitrager
can
borrow one share at the moment $T_i$, sell it for $S_i$ units 
of cash, put
cash in the bank and buy $S_i e^{r\Delta}S^{-1}_{i+1}$ shares at the
moment
$T_{i+2}$. We have an arbitrage situation again. 

We consider the
following quantity
\begin{equation}
   (S^{-1}_i e^{r_2\Delta}S_{i+1} e^{-r_1\Delta}
 + S_i e^{r_1\Delta}S^{-1}_{i+1} e^{-r_2\Delta}
 - 2)/2 \Delta \ .
\label{eq:1}
\end{equation}
It is a sum of excess returns on the plaquette arbitrage operations.
In continuous limit this quantity converges as usual to a 
square of the
curvature tensor element.
The absense of the arbitrage is equivalent to the equality
$$
   S^{-1}_i e^{r_2\Delta}S_{i+1} e^{-r_1\Delta}
 = S_i e^{r_1\Delta}S^{-1}_{i+1} e^{-r_2\Delta} = 1\ ,
$$
and we can use quantity (\ref{eq:1})
to measure the arbitrage (excess rate of return).
In more formal way the expression (\ref{eq:1}) may be written as
$$
R = (U_{1} U_{2} U_{3}^{-1} U_{4}^{-1} + 
U_{3} U_{4} U_{2}^{-1} U_{1}^{-1}
-2)/2 \Delta \ .
$$
In this form it can be generalized for other plaquettes such as, for
example, the "space"-"space" plaquettes.

As we saw above, excess return is an element of the lattice
curvature tensor calculated from the connection. In this 
sense, the rate
of excess return for an elementary arbitrage operation is an 
analogue of
the
electromagnetic field.  In the absence of uncertainty (or, 
in other words,
in the absence of walks of prices, exchange and interest rates) 
the only 
state is realized --- the state of zero arbitrage.  However, if we
introduce
the uncertainty in the game, prices and the rates move and 
some virtual
arbitrage possibilities appear.  Therefore, we can say that 
the
uncertainty plays the same role in the developing theory as 
the  
quantization does for the quantum gauge theory.

Last point to add in the section is a notion of the gauge 
transformation. 
The gauge transformation means a local change of a scale 
in fibres:
$$
f_x\rightarrow g(x) f_x \equiv f_x^{\prime}\ , 
\quad f_x \in F_x \ ,\quad g(x) 
\in G
\ , \quad
x\in E
$$
together with the following transformation of the parallel 
transport operators:
$$
U(y,x) \rightarrow g(y)U(y,x)g^{-1}(x) 
\equiv U^{\prime}(y,x) \in G \ .
$$
It is easy to see that the parallel transport 
operation commutes with a gauge
transformation:
\begin{equation}
g(y)(U(y,x)f_x) = U^{\prime}(y,x)f^{\prime}_x
\label{tr1}
\end{equation}
and the curvature tensor is invariant under the transformation:
\begin{equation}
U_1 U_2 U_3^{-1} U_4^{-1}  = 
U^{\prime}_1 U^{\prime}_2 (U^{\prime}_3)^{-1} 
(U^{\prime}_4)^{-1} \ .
\label{tr2}
\end{equation}

\subsection{Gauge field dynamics}

In the previous subsections we showed that the exchange 
rates and interest rate
(or, more general, internal rate of return) discount factors 
are elements
of the structural group of the fibre bundle. Moreover, they are
responsable for the parallel transport in "space" and 
time directions
correspondingly. In the present subsection we address a 
question about a
dynamics for the exchange/discount factors.

At a first sight the dynamics is difficult to specify 
since it is not  restricted
and any attempt
to formulate the dynamics seems to be voluntary and not enough 
motivated.
However, as we show below the dynamics can be derived from a 
few quite
general and natural ssumptions. 
The main postulate of the present analysis is an assumption about
local gauge invariance with the dilatation group as a gauge group.

{\bf Postulate 1: Gauge invariant dynamics}

We assume that {\it all observable properties of the 
financial enviroment
(in particular, rules of dynamical processes)
do not depend on a choice of units of the assert}s.
It means that all effects of, say, change of 
currency units or shares splittings may
be eliminated by a corresponding change of interest rates, 
exchange rates and
prices~\cite{note42}.
This is a very natural assumption which allow us, however, 
to make a step to a
specification of the dynamics. Indeed, due to the gauge 
invariance an action
which governs the dynamics has to be constructed from gauge 
invariant 
quantities.

{\bf Postulate 2: Locality}
 
Furthermore, we assume {\it local dynamics of the exchange/interest 
rate factors}.
This locality means that in the framework the 
dynamics of an assert is influenced
by connected (in the sense of $\Gamma$ connectivity on the 
base graph $L$) asserts only. 

These two postulates allow us to make the following conclusion: 
the action
$s_{gauge}$ has to be
a sum over plaquettes in the base graph of some function of the 
(gauge
invariant) curvarture of the connection. It means that the 
action is a
sum over plaquettes of a function of the excees return on 
the plaquette
arbitrage operation as it was shown in the previously.

{\bf Postulate 3: Free field theory -- correspondance principle}

We postulate that {\it the action is linear on the plaquette 
curvatures on the base graph} 
since this is a simplest choice.
It will be shown later that the postulate is equivalent 
to the free field
theory description in an absense of matter fields and produce
quasi-Brownian walks for the exchange/interest rates in the 
continuous
time limit. This means that the approach generalizes standard
constructutions of the mathematical finance. This fact 
serves as a
correspondance principle.

{\bf Postulate 4: Extremal action principle}

In fully rational and certain economic enviroment it 
should be no
possibility to have "something from nothing", i.e. to 
have higher return
than the riskless rate of return. In more general form, 
{\it the excess rate of
return (rate of return above the riskless rate) on any kind of 
operations 
takes the smallest possible value} which is allowed by external 
economic
enviroment. Together with the locality of the action this give 
the extremality
principle for the action.

{\bf Postulate 5: Limited rationality and uncertainty}

The real enviroment {\it is not certain and fully rational} and
there exist nonzero probabilities to get other excess rates of
return (exchange rates, prices and interest rates fluctuate and 
its bears local
arbitrage opportunities). 
We assume that the possibilities to have the excess return
$R(x,T)$  at point of "space" $x$ and time moment $T=0$ are
statistically independant for different $x$, $T$ and are
distributed with an exponential probability weight
$e^{-\beta R(x,T)}$ with some effective measure of rationality 
$\beta$.
If $\beta\rightarrow \infty$, we return to a fully rational and
certain economic enviroment. 

Formally, we state that the probability $P(\{U_{i,k}\})$ to
find a set of the exchange rates/ interest rates $\{U_{i,k}\}$ 
is given by the
expression:
$$
P(\{U_{i,k}\}) \sim e^{-\beta \sum_{(x,T)} R(x,T)} 
\sim e^{-\beta s_{gauge}}\ .
$$
Now we are ready to writing down the general action for the 
exchange/interest
rate factors. However, before writing we would like to consider 
in more details
a very simple example which gives some insight of the general 
framework.
Let us, once again, consider two-assert system (cash-shares).

The first moment to mention is a gauge fixing.
Since the action is gauge invariant it is possible to perform a gauge
transformation which will not change the dynamics but will simplify
further calculation. In lattice gauge theory~\cite{Creutz} there are 
several 
standard
choices of the gauge fixing and the axial gauge fixing is one of them.
In the axial gauge an element of the structural group are taken constant
on links in time direction (we keep them $e^{r_{1,2}\Delta}$) and
one of exchange rates (an element along "space" direction at some
particular choosen time) is also fixed. Below we fix the price of the
shares at moment $T=0$ taking $S_0=S(0)$. It means that in the 
situation
of the ladder base the only dynamical variable is the exchange rate 
(price) as
a function of time and the corresponding measure of integration
is the invariant measure $\frac{\mbox{d}S_i}{S_i}$.  

From above derivation,
the definition of the  distribution function for the exchange rate
 (price)
$S=S(T)$ 
at the moment $T=N\Delta$ under condition that at the moment $T=0$ 
the 
exchange rate was $S_0=S(0)$ is given by:
$$
P(0,S_0;T,S) = 
$$
\begin{equation}
\int\limits_0^\infty\!\dots\int\limits_0^\infty\! 
\prod_{i=1}^{N-1}\frac{\mbox{d}S_i}{S_i}
\exp
\left[
 - \frac{\beta}{2 \Delta}\sum_{i=0}^{N-1}
 \left(
   S^{-1}_i e^{r_2\Delta}S_{i+1} e^{-r_1\Delta}
 + S_i e^{r_1\Delta}S^{-1}_{i+1} e^{-r_2\Delta}
 - 2
 \right)
\right] \ .
\label{eq:2}
\end{equation}
It is not difficult to see that in the limit $\Delta \rightarrow 0$ 
the
expression in brackets converges  to the integral
$$
- \beta \int_{0}^{T} d\tau
 \left(
\frac{\partial S(\tau )}{\partial \tau}/S(\tau )   
+ (r_2 - r_1)
\right)^2
$$
which corresponds to the geometrical random walk. Evaluating the
integral
and taking into account the normalization condidtion we come to the
following expression for the distribution function of the price 
$S(T)$:
\begin{equation}
P(S(T)) = \frac{1}{\sigma S \sqrt{2\pi T}}
e^{-(ln(S(T)/S(0)) - (\mu - \frac{1}{2}\sigma^2)T)^2/(2\sigma^2 T)}
\label{log}
\end{equation}
Here we introduced so-called volatility $\sigma$ as
$\beta=\sigma^{-2}$ and the average rate of share return $\mu$ as
$\mu= r_1 - r_2$. 

It is easy to give an interpretation of the last relation.
The system at whole is not concervative and both the interest rate 
$r_1$ 
and the rate $r_2$ come outside of the system (from banks and the 
corresponding company production). Let us imagine that the world 
is certain and 
that because of the company
production the capitalization of the firm increased. For this amount new 
shares with the same price $S_1$ have been ussued (no dividends have 
been payed). 
The number of new shares for each old share is equal $e^{r_2 \Delta}$. 
It means that the cumulative (old) share will have a price 
$S_1 e^{r_2 \Delta}$ while the original price (at zero moment) was 
$S_0$.
Taking into account discounting and certainty, we end with 
the following expression:
$$
S_1 = e^{(r_1 - r_2) \Delta} S_0 \ ,
$$
which tells that the rate of return on the share is equal $r_1-r_2$.
After introducing an uncertainty last expression turns into an average 
rate of return on the share.

Eqn (\ref{log}) returns us to a justification of Postulate 3 
which, together with other Postulates,
are equivalent to a log-normal model for a price walks in an absence
of matter fields, which we consider in details in the next section. 

Now we give the general expression for probability distribution of a 
given
exchange rates and internal rates of return profile:
\begin{equation}
P(\{S_{i,k} \},\{r_{i,k}\}) \sim
exp( -\frac{\beta}{\Delta} 
\sum_{plaquettes} (\prod_{m} U_{m} + 
\prod_{m} U_{m}^{-1} - 2)) \ ,
\end{equation}
where the sum is calculated over all plaquettes in the base graph with
all links with nonzero connectivity $\Gamma$, the index $m$ runs over all 
links of a plaquette, $\{U_{m}\}$ are corresponding elements of the
structural group, which perform the parallel transport along the links. 

To complete the gauge field consideration we want to return once
again to the gauge invariance principle.
Next section is devoted to the consideration of the ``matter" field
which interact through the connection. It is clear now that the field is
a field which represents a number of assets (as a phase in the 
electrodynamics)
and has to be gauged by the connection.
Dilatations of money units (which do not change rules of investors
behaviour) play a
role of gauge transformation which eliminates the effect of the
dilatation by a proper tune of the connection (interest rate, exchange
rates, prices and so on) exactly as it is in the Fisher formula for the
real interest rate in the case of inflation~\cite{Lumby}.  The symmetry
to a local dilatation of money units (security splits
and the like) is the gauge symmetry of the theory.

\subsection{Effective theory of cash-debts flows: matter fields}
 
Now let us turn  our attention to "matter" fields. These fields 
represent cash-debts flows on the market. Importance of the cash-debts
flows for our consideration is caused by their role in a stabilization 
of 
market prices. Indeed, if, say, some bond prices eventually 
go down and create a
possibility to get bigger return than from other assets, then an
effective cash flows appear, directed to these more valuable bonds.
This causes an upward shift of the prices because of the demand-supply
mechanism. All together these effects smoothen  the price movements.  
The same picture is valid for debts flows if there is a possibility for
debts restructurisation. As we will see all this features finds their
place in the constructed framework.  
 
To formulate an effective theory for the flows we
will assume that:

\begin{enumerate}
\item Any particular investor tries to maximize his return on cash 
and
minimize his debts.
\item Investor's behaviour is {\it limited} rational i.e. there are
deviations from pure rational strategy because of, for example,
lack of complete information, specific financial manager's objectives
\cite{Lumby} and so on.
\end{enumerate}

We start with a construction of an effective theory for the cash
flows and the generalize it for the case of debts presence.
The first assumption tells us that an investor tries to maximize the
following
expression for the multiplier of a value of his investment 
(in the case of
cash, shares and securities):
$$
s(C) = ln( U_1 U_2 .... U_N )/ \Delta
$$
by a certain choice of the strategy which is resulted in the
corresponding trajectory in "space"-time for assets.
Here $\{U_i\}_{i=i}^{N}$ are exchange (price) or interest factors 
which
came from a
choice of the investor behaviour at $i$-th step on the trajectory 
$C$ and
boundary points (at times $T=0$ and $T=N$) are
fixed. We assume that there is a transaction time which is
smallest time in the systems  equal to $\Delta$.
In other words, a rational investor will choose the trajectory
$C_0$ such that
$$
s(C_0) = max_{\{C\}} s(C) \ .
$$

Choosing the best strategy, fully rational investor maximizes
his return  $s$. However, as we assumed limited rationality, in analogy
with the corresponding consideration of the connection field 
probability
weights, we define the following probability weight for a certain
trajectory $C$ with $N$ steps:
\begin{equation}
P(C) \sim e^{\beta^{\prime} s(C)/\Delta } \equiv \ e^{\beta s(C)},
\label{pc}
\end{equation}
with a some "effective temperature" $1/ \beta$ which represents a
measure of average unrationality of investors in a unit time.

It is possible to generalize the approach to a case of many 
investors operating with cash and debts. The
corresponding functional integral representation for the transition 
probability (up to a normalization constant) has the form~\cite{KI}:
$$
\int D\psi_{1}^+ D\psi_1 D\psi_{2}^+ D\psi_{2} D\chi_{1}^+
D\chi_1 D\chi_{2}^+ D\chi_2 e^{\beta (s1 + s1^{\prime})} \ ,
$$
with the actions for cash and debts flows:
$$
s1= \frac{1}{\beta} \sum_{i} (
\psi^{+}_{1,i+1} e^{\beta r_1 \Delta} \psi_{1,i} - 
\psi^+_{1,i}\psi_{1,i}
+
\psi^{+}_{2,i+1} e^{\beta r_2 \Delta} \psi_{2,i} - 
\psi^+_{2,i}\psi_{2,i}
$$
\begin{equation}
+ (1-t)^{\beta}S_i^{\beta}  \psi^+_{1,i+1} \psi_{2,i}
+ (1-t)^{\beta} S_i^{-\beta} \psi_{2,i+i}^+ \psi_{1,i}) \ ,
\label{s1}
\end{equation}        
$$
s1^{\prime} = \frac{1}{\beta} \sum_i (\chi_{1,i+1}^+ 
e^{-\beta r_1 \Delta}
\chi_{1,i}
- \chi^+_{1,i}\chi_{1,i} + 
\chi_{2,i+1}^+ e^{-\beta r_2 \Delta} \chi_{2,i}
$$
\begin{equation}
- \chi^+_{2,i} \chi_{2,i} + 
(1+t)^{-\beta}S_i^{-\beta}  \chi^+_{1,i+1} \chi_{2,i}
+ (1+t)^{-\beta}S_i^{\beta} \chi_{2,i+1}^+ \chi_{1,i} ) \ .
\label{s1'}
\end{equation}
We do not stop here to describe the gauge invariant boundary 
conditions which
are discussed in details in Refs~\cite{KI,ISfin1}.

It is interesting to note that the Eqn(\ref{s1'}) may be transformed 
to
Eqn(\ref{s1}) inverting signs of $r_i$ and inverting exchange rates 
$S$ in the
absense of the transaction costs. This corresponds to the
fransformation from negative to a positive charge. The transaction 
costs 
make this symmetry only approximate. 

In the absence of restructuring of debts (i.e. one kinds of debts 
cannot
be transfromed to another kind of debts) last terms containg $S$ 
have to
be dropped out. Then the action takes esspecially simple form:
\begin{equation}
s1^{\prime}_0 = 
\frac{1}{\beta}
\sum_i (\chi_{1,i+1}^+ e^{-\beta r_1 \Delta} \chi_{1,i} - 
\chi^+_{1,i}\chi_{1,i}
+
\chi_{2,i+1}^+ e^{-\beta r_2 \Delta} \chi_{2,i} - 
\chi^+_{2,i}\chi_{2,i} )
\ .
\label{s1'0}
\end{equation}
  
All said above was related to the particular case of two assets problem.
Now we give a form of the action for the most general situation. As 
it was shown
before, a general system of assets is described by a form of the 
"space"-time
base graph $L$ of the fibre bundle. Elements of the graph are labeled   
by $(i,k)$ where the index $i$, $i=0,..,N$, represents a particular 
asset (a point in
"space") and the index $k$ labels time intervals. By the definition 
of the base
$L$ any two points of the base can be linked by a curve $\gamma$, each
formed by elementary segments with a nonzero connectivity $\Gamma$.
In its own, each of the segments is provided with an element of 
the structural
group $U$ which performs a parallel transport along this (directed) link.
These allow us to give a most general form of the action $s1$
($(i_{12},k_{12})\equiv (i_1,k_1),(i_2,k_2)\in L  
:\Gamma ((i_1,k_1),(i_2,k_2))=1$):
\begin{equation}
s1= \frac{1}{\beta} 
\sum_{(i_{12},k_{12})}
( \psi_{(i_1,k_1)}^+  U^{\beta}_{(i_12,k_12)} (1 - t
\delta_{i_1-1,i_2})^{\beta}
\psi_{(i_2,k_2)} - 
\delta_{(i_1,k_1),(i_2,k_2)} \psi^+_{(i_1,k_1)} \psi_{(i_1,k_1)}) \ .
\nonumber
\end{equation}
In the same way the action for the debts flows can be written:
\begin{equation}
s1^{\prime} = \frac{1}{\beta}
\sum_{(i_{12},k_{12})}
( \chi_{(i_1,k_1)}^+ U^{-\beta}_{(i_12,k_12)} (1 + t
\delta_{i_1-1,i_2})^{-\beta}
\chi_{(i_2,k_2)} -
\delta_{(i_1,k_1),(i_2,k_2)} \chi_{(i_1,k_1)}^+ \chi_{(i_1,k_1)}) \ .
\nonumber
\end{equation}

\section{Conclusion}

In conclusion, we proposed a mapping of the capital market 
theory in the 
lattice
quantum gauge theory where the gauge field represents the 
interest rate and 
prices and
"matter" fields are cash-debts flows. Basing on the mapping 
we derived 
action
functionals for both the gauge field and "matter" fields 
assuming several
postulates.  The main assumption is a gauge invariance of 
the dynamics which 
means
that the dynamics does not depend on particular values of units 
of assets and a
change of the values of the units may be compensated by a proper 
change of the 
gauge
field. The developed formalism has been applied to some issues of 
the capital market theory. 
Thus, it was shown in Ref~\cite{ISfin1} that a deviation of the 
distribution function from the log-normal
distribution may be explained by an active trading behavior
of arbitragers. In this framework such effects as changes
in shape of the distribution function,
``screening" of its wings for large values of the price and 
non-Markovian   
character (memory) of price random walks turned out to be 
consequences of
damping of the arbitrage and directed price movements caused by
speculators.
In Ref~\cite{II1} consequences of the bid-ask spread and 
the corresponding gauge invariance breaking were examined. 
In particular, it turned out that the
distribution function is also influenced by the bid-ask spread 
and the change of its
form may be explained by this factor as well. So, the complete 
analysis of the 
statistical characteristics of prices have to account both 
these factors.
Besides, it is possible to show~\cite{IK1} that the 
Black-Scholes equation 
for the financial derivatives
can be obtained in the present formalism in an absence 
of speculators (i.e. 
absence of the arbitrage game). 

Let us now make two final remarks.
\begin{enumerate}
\item
In all cited above references a very simple model of a stock 
exchange was 
considered, where the only
one kind of security is traded. The consideration can be 
generalized on more
realistic situation with a set of traded securities. Following this 
line,
dynamical portfolio theory can be constructed and, in static 
(equilibrium)  
limit, will coincide with standard portfolio 
theory~\cite{portfolio}.
In the dynamical theory time-dependent correlation functions will
play a role of response functions of the market to an external
perturbation
such as a new information or a change in macroeconomic environment.
Account of virtual arbitrage fluctuations will lead to a 
time-dependent
modification of CAPM~\cite{Blake}.
\item
Since the influence of the speculators leads to a non-Markoffian 
character of a
price walks there is no possibility
to eliminate a risk using arbitrage arguments to derive an 
equation for  
a price of a derivative. 
Then the virtual arbitrage and corresponding asset
flows
have to be considered. It will lead to a
correction of Black-Scholes equation.
\end{enumerate}
We will return to these points in our forthcoming papers.

\section*{Acknowledgment}
Author wants to thank Alexandra Ilinskaia and Alexander Stepanenko for
useful discussions. The work was supported by UK EPSRC grant 
GR/L29156 and by 
grant RFBR N 95-01-00548.


\begin{thebibliography}{99}

\bibitem{KI}
K. Ilinski : {\it Gauge Theory of Arbitrage}, 
Iphys Group working paper IPHYS-1-97, 1997;

\bibitem{Lumby} S. Lumby, {\it Investment appraisal and financial
decisions}, Chapman $\&$ Hall, 1994;

\bibitem{Blake} D. Blake, {\it Financial Market Analysis},
McGraw-Hill, 1990;

\bibitem{Wilmott} P. Wilmott, S. Howison and J. Dewynne,
{\it The Mathematics of Financial Derivatives}, Cambridge University
Press, 1995;

\bibitem{Duffie} D. Duffie,
{\it Dynamic Asset Pricing Theory}, Princeton University
Press, 1992;

\bibitem{note1}
It is possible to pose such capital market
various imperfections and restrictions for lending and borrowing (for
example, certain credit limit). This leads to a effective quantum 
systems
with constraints. 

\bibitem{EGH} T.Eguchi, P.B. Gilkey and A.J. Hanson:
Gravitation, gauge theories and differential geometry,
{\it Physics Reports}, {\bf 66} N6, 213 (1980);
B.A. Dubrovin, A.T. Fomenko, S.P. Novikov: "Modern Geometry --
Methods and Applications", Springer-Verlag, 1984;  

\bibitem{note42}
In practice some violation of the dilatation invariance appear 
because of
a some structures of transaction costs and bid-ask spread 
(I am grateful
to P.A.Bares for this comment);

\bibitem{Creutz} M. Creutz, {\it Quarks, gluons and lattices},
Cambridge University Press, 1983; 

\bibitem{ISfin1}
K. Ilinski and A. Stepanenko: {\it How arbitragers change log-normal
distribution}, preprint IPhys-2-97, 1997;

\bibitem{II1}
A. Ilinskaia and K. Ilinski: {\it Bid-ask spread as a gauge invariance 
breaking}, preprint IPhys-4-97, 1997;

\bibitem{IK1}
K. Ilinski and G. Kalinin: {\it Black-Scholes equation from Gauge 
Theory of Arbitrage}, in preparation;

\bibitem{portfolio} E.J. Elton, M.J. Gruber,
{\it Modern portfolio theory and investment analysis}, Jonh Wiley 
$\&$ Sons, 1995;
\end{thebibliography}
\end{document}